%% file: 15587.tex
\documentclass[]{aa}

\usepackage{txfonts}
\usepackage{graphicx}
\usepackage{natbib}
\usepackage{longtable}

\begin{document}

\title{Photospheric and coronal abundances in solar-type stars:\\
the peculiar case of \object{$\tau$\,Bootis}}

\author{A.~Maggio\inst{\ref{i1}} \and J.~Sanz-Forcada\inst{\ref{i2}} \and
L.~Scelsi\inst{\ref{i1}}}
\offprints{A.~Maggio, {\email maggio@astropa.inaf.it}}
\institute{INAF - Osservatorio Astronomico di Palermo, Piazza del
Parlamento 1, 90134 Palermo, Italy, \email{maggio@astropa.inaf.it,
scelsi@astropa.inaf.it}\label{i1} 
\and 
Centro de Astrobiolog\'ia,
CSIC-INTA, European Space Astronomy Center, P.O. Box 78, E-28691, Villanueva de
la Ca\~nada, Madrid, Spain, \email{jsanz@laeff.inta.es}\label{i2}
}

\date{14 August 2010 / 29 November 2010}

\titlerunning{Photospheric and coronal abundances of $\tau$\,Bootis}
\authorrunning{A.~Maggio et al.}

\abstract
{}
{Chemical abundances in solar-type stars are still a much debated topic
in many respects. In particular, planet-hosting stars are known to be
metal-rich, but whether or not this peculiarity applies also to the
chemical composition of the outer stellar atmospheres is still to be
clarified. More in general, coronal and photospheric abundances in
late-type stars appear to be different in many cases, but understanding
how chemical stratification effects work in stellar atmospheres requires
an observational base larger than currently available.}
{We obtained XMM-Newton high-resolution X-ray spectra of $\tau$\,Bootis, 
a well known nearby star with a Jovian-mass close-in planet. We analyzed
these data with the aim to perform a detailed line-based emission
measure analysis and derive the abundances of individual elements
in the corona with two different methods applied independently. 
We have compared the coronal abundances of $\tau$\,Bootis with published
photospheric abundances based on high-resolution optical spectra and
with those of other late-type stars with different magnetic activity
levels, including the Sun.}
{We find that the two methods provide consistent results within the
statistical uncertainties for both the emission measure distribution of
the hot plasma and for the coronal abundances, with
discrepancies at the $2\sigma$ level limited to the amount of plasma at
temperatures of 3-4 MK and to the O and Ni abundances. In both cases, 
the elements for which both coronal and photospheric measurements are
available (C, N, O, Si, Fe, and Ni) result systematically less abundant 
in the corona by a factor 3 or more, with the exception of the coronal Ni
abundance which is similar to the photospheric
value. Comparison with other late-type stars of similar activity level
shows that these coronal/photospheric abundance ratios are peculiar to
$\tau$\,Bootis and possibly related to the characteristic over-metallicity of this
planet-hosting star.}
{}

\keywords{stars: late-type - stars: atmospheres - stars: coronae - stars: individual:
$\tau$\,Bootis - stars: abundances - X-rays: stars}

\maketitle

\section{Introduction}

Chemical abundances in solar-type stars are still a much debated topic.
Solar abundances have been derived both from optical/UV/X-ray
spectroscopy and from composition measurements of solar wind particles or meteorites
\citep{fl00}, but the most recent and sophisticated results are in stark conflict
with models of the solar interior tuned with helioseismology data \citep{agss09}.
A second problematic point is the finding that photospheres of stars hosting close-in giant
planets are known to be metal-rich, which can possibly be explained by a higher
probability of planet formation in high-metallicity birth clouds \citep{gonzalez06}. 

Another conundrum is offered by the differences found in the Sun and other
solar-like stars between photospheric and coronal abundances. In this respect,
stellar X-ray spectroscopy is a fundamental tool for the chemical analysis of stellar
outer atmospheres, and the only method to determine the abundances of noble gases, 
like argon and neon, with no optical lines in photospheric spectra.
For the solar corona, and in particular in long-lived coronal structures,
the composition of the plasma appears enriched with elements having low first ionization
potentials (FIP $< 10$\,eV) by about 
a factor 4, on average, with respect to photospheric values \citep{fl00}. 
In other stars a more complex behavior has been observed 
\citep{gn09}, with a tendency for the low-FIP elements (including
iron) to become depleted with respect to the high-FIP elements (neon
in particular) in extremely active RS CVn-type and Algol-type binaries.
Several theoretical explanations have been proposed, but our understanding is still
largely driven by the observations, which indicate a dependence of the
coronal/photospheric abundance ratios on the stellar magnetic activity level
\citep{rsf08} or on the spectral type \citep{wl10}. 

The aim of the present work is to investigate the coronal/photospheric abundance problem
in a metal-rich planet-hosting star. To this aim we made use of detailed
analyses of the photospheric composition, available in the literature, for stars of this
class, and of X-ray data acquired on purpose with the XMM-Newton satellite.  
The comparison of our target with other solar-type stars with similar levels 
of magnetic activity provides new information on the relative importance of the photospheric
composition in the abundance stratification mechanism(s) operating in stellar
atmospheres. In Sect.\ \ref{sec:targ} we present our sample star, and in Sect.\
\ref{sec:photab} its photospheric chemical composition. The analysis of the X-ray
spectra, including the derivation of the plasma emission measure distribution (EMD) vs.\
temperature and coronal abundances is described in Sect.\ \ref{sec:xray}. Sections
\ref{sec:discuss} and \ref{sec:concl} are devoted to discussion and
conclusions.

\section{Why $\tau$\,Bootis}
\label{sec:targ}

$\tau$\,Bootis (\object{HR 5185}, \object{HD 120136}) was selected as part of a project aimed 
at studying the
coronal abundances of late-type dwarfs classified as super-metal-rich
stars, and sufficiently X-ray bright to obtain good X-ray spectra
with the XMM-Newton Reflection Grating Spectrometers (RGS1 and RGS2).

$\tau$\,Boo is a $1.3\,{\rm M_{\sun}}$ F7 V star
\citep{perrin+77,baines+08}, located at 15.6\,pc from the
Sun. It harbors a planet with a mass $m \sin i \approx 3.9\,{\rm M_{\rm J}}$,
discovered by \citet{butleretal97} in a 3.31 day period.  The stellar
age was estimated by several authors with different techniques (in
particular, position in the H-R diagram and chromospheric emission level)
and constrained within the range 1--3 Gyr \citep{takedaetal07}.
However, based on its X-ray luminosity \citep[$L_{\rm x} \approx
10^{29}$\,erg s$^{-1}$, ][]{schmitt+85}, the age of $\tau$\,Boo could be as young
as 0.4\,Gyr \citep{jsf+10}.

The rotation characteristics of $\tau$\,Boo were determined by
\citet{reiners06}, \citet{catalaetal07}, and \citet{donatietal08}.
The photospheric line profiles are reasonably well
fitted assuming an homogeneous surface rotational velocity $v \sin i =
15 \pm 1\,{\rm km s^{-1}}$ and a turbulent velocity of 5.5\,km
s$^{-1}$. Fourier analysis of the same data
was employed to establish a surface differential
rotation of $\sim 20$\%, which implies a rotation period at the equator of
3.0\,d, and 3.7--3.9\,d at the pole, assuming a rotation axis
inclined at an angle $i = 40^{\circ}$ with respect to the line
of sight. Hence, $\tau$\,Boo is considered the first convincing case of a
star whose rotation (at intermediate latitudes) is synchronized with 
the orbital motion of its close-in giant planet.  

\citet{catalaetal07} and \citet{donatietal08}
also studied the strength, topology, and time change of the surface magnetic 
field of $\tau$\,Boo. At large spatial scales, the field comprises a
dominant poloidal component, more complex than a dipole, and a small
toroidal component. Field strengths of up to 10 G were found, and the
overall polarity turned out to be reversed between two observations 
taken about one year apart. Most recently, a second polarity reversal 
was reported by \citet{faresetal09}, suggesting a magnetic cycle of about two
years.

Finally, $\tau$\,Boo is also one of the stars recently investigated to
search for evidence of star-planet interaction effects.
\citet{shkolniketal08} and \citet{walkeretal08} presented
studies of the photospheric (white light) and chromospheric (\ion{Ca}{ii}
K line core) variability of $\tau$\,Boo over a time range of a few years, 
suggesting the presence of a persistent dark spot and enhanced
chromospheric emission synchronized with the planet orbital period, but
occurring at phase $\phi \sim 0.8$ (i.e. preceding the sub-planetary
stellar longitude, $\phi = 0$, by $\sim 70^{\circ}$). These results are
suggestive of variability caused by an active region induced by a
star-planet magnetic interaction, but the signal
appears marginally significant, and further studies are required to
confirm this conclusion \citep[see also][]{faresetal09}.

In summary, $\tau$\,Boo is a relatively massive star with a shallow
convection zone and a fairly massive planet in a tight orbit. The
magnetic activity of $\tau$\,Boo is moderate and typical of mid-F dwarfs,
with little if any modulation at either the orbital or rotation period.
The near synchronization between the star's rotation and the planetary
orbit might imply little or no stress of the magnetic field lines that
connect the two components of the systems, and hence a low
planet-induced activity enhancement. In this respect, $\tau$\,Boo is an
ideal target for studying element abundances in the quiescent corona of
a star with a photospheric metallicity significantly exceeding the solar
one.

\section{Photospheric abundances}
\label{sec:photab}

The composition of Tao~Boo's photosphere was determined by several
authors in the context of many studies carried out to quantify and
explain the peculiar metallicities of stars with planets
\citep[SWPs; see][for a review]{gonzalez06}. 

We selected $\tau$\,Boo originally from the extensive compilation of
\citet{tay94c}, who has critically examined the
high-resolution spectroscopic abundance determinations of F, G, and K 
stars available in literature.  $\tau$\,Boo was noted as a
metal-rich star according to the criterion [Fe/H] $>~0.2$.
Subsequently, the high metallicity of our target was confirmed 
by several accurate abundance analyses, the most recent performed by
\citet{ecuvillonetal04}, \citet{valentifischer05}, \citet{takedahonda05}, 
\citet{gillietal06}, \citet{luckheiter06}, and \citet{takeda07}.

Most of the above studies were considered in the work by
\citet{gonzalezlaws07}, who have presented the chemical abundances for
18 elements in 31 SWPs and compared them with a sample of nearby stars
without detected planets. In order to cope with differences in the
zero-point, effective temperature and abundance scales among the various
studies, these authors have
applied statistical corrections to abundance measurements for individual
elements reported by each author. For these corrections, they set 
\citet{gillietal06} as the reference\footnote{In turn, the 
stellar abundances
in \citet{gillietal06} are give with respect to the solar photospheric abundances
by \citet{ag89}.} for all the elements, 
except for C and O, for which \citet{ecuvillonetal04} and \citet{luckheiter06}, 
respectively, were adopted. Finally, they computed average
abundance values for each element, and standard deviations from
individual uncertainties summed in quadrature. Note that this procedure
yields conservative (slightly overestimated) uncertainties 
on the final values, larger than the scatter between individual 
measurements. The abundance values of $\tau$\,Boo computed
by \citet{gonzalezlaws07} are reported in
Table~\ref{tab:photab}, where we also added the single nitrogen abundance
determined by \citet{ecuvillonetal04b}. These are the photospheric abundances 
we adopted as reference for comparison with the coronal abundances 
derived by us.

\input{15587_tab1}
\input{15587_tab2}

\section{X-ray emission}
\label{sec:xray}

\begin{figure} 
\centering
\resizebox{\hsize}{!}{\includegraphics[angle=90.]{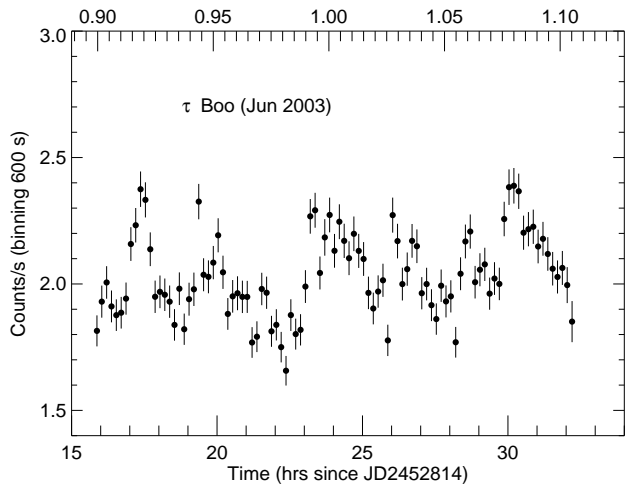}}
\caption{X-ray light curve of $\tau$\,boo obtained with EPIC (summed PN and MOS broad-band
photometry).  Time bins are 600\,s long. The upper X-axis scales indicate
the planetary orbital phase at the time of the observation.}
\label{fig:lc}
\end{figure} 

$\tau$\,Boo was observed on June 24, 2003, for a duration of about 65\,ks, with
the {\sc thick} filter in front of the EPIC detectors, the MOS in {\sc
large window} mode, and the PN in {\sc full frame} mode.
After screening for high-background intervals, the useful
exposure time was reduced to about 50\,ks, during which the source yielded a
mean count rate of 0.3\,s$^{-1}$ in the MOS, and 1.3\,s$^{-1}$ in the
PN (0.3 -- 5\,keV band).

Figure~\ref{fig:lc} shows the X-ray light curve of $\tau$\,Boo with a bin
size of 600\,s, just characterized by some low-level variability,
typical of solar-type stars with moderate activity.
According to the ephemeris by \citet{catalaetal07}
\begin{equation}
T = {\rm HJD}~2453450.984 + 3.31245~E
\end{equation}
during the 16 hours of the XMM-Newton observation the planet orbital phase was
between 0.89 and 1.13, i.e. the planet was near the
opposition\footnote{Note that the zero-point phase in
\citet{catalaetal07} is shifted by half a period with respect to the
ephemeris in \citet{walkeretal08} and in \citet{shkolniketal08} (based
on older photometric data), where phase 0 is with the planet in front
of the star.} (occulted by the star).

The X-ray spectral analysis was carried out in several steps:
first, a preliminary fit to the EPIC data with three-component thermal
models was performed to determine the level of continuum emission and to
guess the plasma metallicity; these results were employed to proceed
with the measurement of emission lines in the RGS high-resolution
spectra, and eventually for a detailed line-based emission
measure analysis, including an estimate of chemical abundances for a
number of elements with measurable spectral signatures (in particular, 
C, N, O, Ne, Fe, and Ni); the plasma emission measure distribution and 
available element abundances were employed subsequently to determine
high-temperature components and other element abundances (Mg, Si, S, Ar)
by again fitting the EPIC spectra. Below we explain these
steps more in detail.
 
\begin{figure*} 
\centering
\resizebox{\hsize}{!}{\includegraphics[]{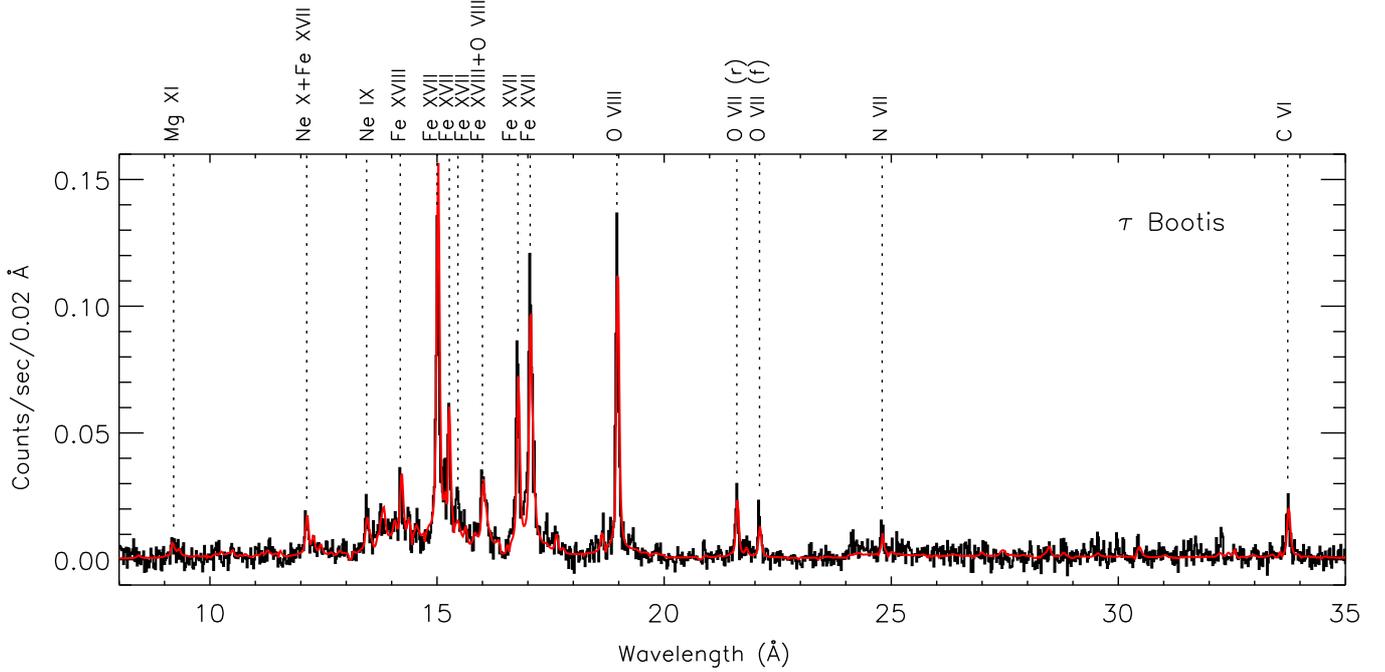}}
\caption{Summed RGS1 and RGS2 spectra of $\tau$\,Bootis (histogram) with
the synthetic model spectrum derived from the analysis results obtained
with Method 1. Labels identify some of the strongest lines.} 
\label{fig:rgsspec}
\end{figure*} 

\subsection{RGS data analysis}

The background-subtracted RGS1 and RGS2 spectra of $\tau$\,Boo contain
$\sim4000$ and $\sim5100$ net counts, respectively (Fig.~\ref{fig:rgsspec}).

The analysis of X-ray emission line spectra is notoriously difficult,
and the measurement of element abundances can be affected by systematic
uncertainties that are difficult to quantify because of the assumed plasma
emissivity model, the identification and measurement of relevant lines,
and the algorithm adopted for the emission measure analysis
\citep{maggiobexst}. For this reason, we employed two different
approaches for the analysis of the XMM-Newton grating spectra of
$\tau$\,Boo, carried out in a completely independent manner.

\subsubsection{Method 1}
\label{sec:emd1}

The first approach is the one discussed in detail in \citet{scelsi04} together
with a study of its accuracy; here we limit ourselves to report the main points
of this iterative method.

We employed the software package PINTofALE V2.0 \citep{PoA2000} and, in part, 
also XSPEC, and used the Astrophysical Plasma Emission Database
\citep[APED/ATOMDB V1.3,][]{aped2001}, which includes the
\citet{Mazzotta1998} ionization equilibrium.

Background-subtracted RGS1 and RGS2 spectra were first re-binned by a factor of
2 and then co-added for the identification of the strongest emission lines
and the measurement of their fluxes. In the rebinned RGS1+RGS2 spectrum we
identified about 40 lines from \ion{Fe}{xvii-xxi}, \ion{Ne}{ix-x}, 
\ion{O}{vii-viii}, \ion{N}{vii}, \ion{C}{vi}, and \ion{Ni}{xix-xx} ions.
After this step the method proceeds in an iterative way.

To measure line fluxes we adopted a Lorentzian line profile and
initially assumed the continuum level evaluated from a 3-T model 
best-fitting the {\sc pn} spectrum. Indeed, the wide line wings make it impossible
to determine the true source continuum below $\sim 17$\,\AA\ directly from the
RGS data, especially in the $\sim 10-17$\,\AA\ range where the spectrum is
dominated by many strong overlapping lines. The \ion{Mg}{xi} triplet was
measured as a single feature by counting the photons within the
$9.10-9.40$\,\AA\ range in the RGS spectrum rebinned by a factor of 5, and
subtracting the assumed continuum contribution in the same wavelength region.
We also fixed $N_{\rm H}=3\times 10^{18}$\,cm$^{-2}$, compatible with the short
distance of $\tau$ Boo (15.6\,pc).
The results are listed in Table~\ref{tab:lines}. 

To reconstruct the emission measure distribution (EMD,
defined as $E\!M(T) = \int_{\Delta T} n_H n_e dV$ [cm$^{-3}$])
and to estimate element abundances, we selected a set of 
19 lines among those identified that had reliable flux measurements
and theoretical emissivities. 
Most of them are blended with other lines, so the measured
line flux is actually the sum of the contributions of a number of atomic
transitions. Accordingly, the ''effective emissivity'' of each
line blend must be evaluated as the sum of the emissivities of the lines 
that mostly contribute to that spectral feature. 
If the blend includes lines from different
elements, the effective emissivity needs to be weighted by their relative
abundances. Since these abundances are not available at the beginning 
of the analysis, we initially selected 
only iron lines not blended with lines of other elements, and the latter
are included one at a time in the following stages.

We performed the EMD reconstruction with the Markov-Chain Monte Carlo (MCMC)
method by \citet{KashyapDrake1998}. This method yields a volume emission measure
distribution, $EM(T_{k})=dem(T_{k})\,\Delta \log T$, and related statistical
uncertainties $\Delta EM(T_k)$, where
$dem(T)=n_{\rm e}^{2}\,{\rm d}V/{\rm d} \log T$ 
is the differential emission measure of an optically thin plasma 
and $\Delta \log T =0.1$ is a constant bin size\footnote{We also performed
the EMD analysis with a coarser $\Delta \log T =0.2$ temperature grid,
but obtained consistent results within statistical uncertainties.}. 
The method also provides estimates of element abundances scaled by the
iron abundance, with their statistical uncertainties. The iron abundance is estimated by
computing the emission measure distribution assuming different metallicities
and by comparing the synthetic continuum with the observed one at
$\lambda >20$\,\AA\ in the RGS spectrum, which is a spectral region free of
strong overlapping emission lines. Finally, we checked the solution obtained
with the MCMC by comparing (i) the line fluxes predicted from our solution
with the measured ones and (ii) the whole model spectrum with the observed one
(Fig. \ref{fig:rgsspec}).

The procedure described above is repeated two or 
three times to check line identifications and to ensure consistency between 
the continuum level assumed for flux
measurements and the continuum predicted by the EMD. Indeed, the continuum is
adjusted at each iteration because it may become
different from that predicted by the $3-T$ model best-fitting the {\sc pn}
spectrum, which is adopted as the initial guess. This procedure
ensures that possible cross-calibration offsets between {\sc pn} and RGS do
not affect the final $EMD$. 

\begin{figure} 
\resizebox{\hsize}{!}{\includegraphics{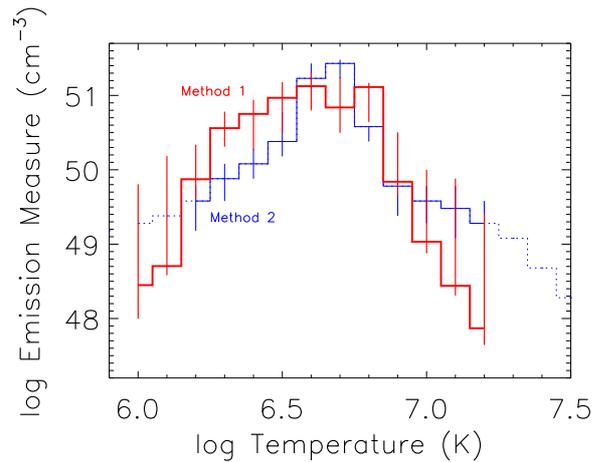}}
\caption{Emission measure distribution of $\tau$~Boo based on the XMM-Newton
RGS spectra. The thick (red) histogram was obtained with Method 1 \citep{scelsi04}, the thin
(blue) histogram with Method 2 \citep{sanz03}. In the latter, dot-lined bins are more uncertain 
(no associated error bar).
}
\label{fig:emd} 
\end{figure}

The final $EM(T)$ values are reported in Table~\ref{tab:emd} 
with the coronal abundances of individual elements. The EMD is also shown in
Fig.\ \ref{fig:emd} with the distribution obtained with Method 2.

\input{15587_tab3}

\subsubsection{Method 2}
\label{sec:emd2}

The second approach for constructing the EMD vs.\ temperature
follows the line-based analysis described in \citet{sanz03} and references
therein. In short,
individual line fluxes are measured and those with a signal-to-noise
ratio $S/N >$ 3--4 are then compared with the theoretical values obtained from a
trial EMD, combined with the same atomic emission model
used for Method 1. The comparison of
measured and model line fluxes results in an improved EMD that is
used again to produce new model line fluxes. 

In Method 2 we employed 27 lines. The first iteration is based only 
on Fe lines, either blended or unblended, assuming solar abundances. 
The successive
iterations are used to calculate the abundances of the different
elements, that are required to refine in particular the line predictions for 
all the blends. 
Because the lines corresponding to the various elements are formed
in different temperature ranges, we take the caution to
progressively add lines of elements with formation temperatures 
that overlap the EMD computed at each iteration. 

The method converges in a few iterations, and is
therefore self-consistent.  This process yields
an EMD solution that is not unique, but reliably matches the
model fluxes to the observed ones, and it presumably describes the
actual amount of plasma at different temperatures in corona. 

Finally, error bars are calculated using a Monte Carlo method
that seeks the best solution for different line fluxes within their
1--$\sigma$ errors \citep[see][ for more details]{sanz03}. 

In summary, the two methods differ in the choice of the line set employed 
for the EMD reconstruction, in the algorithm to search for the best
EMD solution, and in the evaluation of its confidence range. 

\begin{figure} 
\resizebox{\hsize}{!}{\includegraphics[angle=270.]{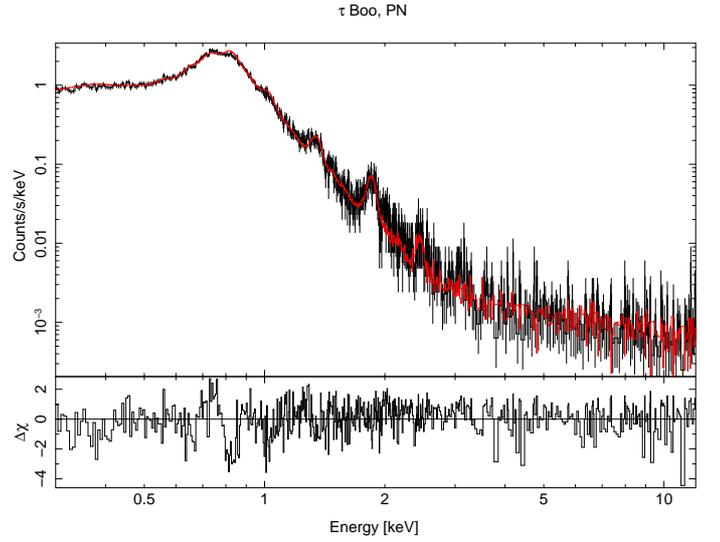}}
\caption{{\it Upper panel:} PN observed spectrum (black histogram with error bars)) 
and best-fit model (in red). {\it Lower panel:} residuals in sigma units.}.
\label{fig:epicspec} 
\end{figure} 

\subsection{EPIC spectral analysis}

Typical EPIC spectra can usually be fitted with thermal models with few (2 -- 4)
isothermal components. This is an approximation to the actual emission
measure distribution of the coronal plasma, which provides a reasonably
good quality of the fit to individual CCD-resolution X-ray
spectra (MOS or PN); instead, systematic errors and cross-calibration uncertainties 
make the joint analysis of all the available spectra (EPIC and RGS) cumbersome,
and yield best-fit $\chi^2$ values formally too high to be statistically
acceptable. This is also the case for $\tau$~Boo.

The simplest multi-component thermal model that provides an acceptable fit
to the EPIC spectra is a three-temperature model with plasma at 2\,MK, 4\,MK, and 8\,MK
and a volume emission measure in the proportions 1:6:1 (see Table \ref{tab:emd} and Fig.\
\ref{fig:epicspec}). This model fitting confirms the low interstellar absorption assumed in the
analysis of the RGS spectra (Sect. \ref{sec:emd1}), and yields an independent measure of the
abundances of O, Ne, Mg, Si, and Fe. Note that Mg and Si abundances can be effectively constrained 
by this fitting, because the EPIC spectra clearly show line complexes caused by
\ion{Mg}{xi} (1.33--1.35\,keV), \ion{Mg}{xii} (1.47\,keV), 
and \ion{Si}{xiii} (1.84--1.86\,keV).  
Abundances of other elements cannot be constrained by
the EPIC spectra, and they have been fixed either to the values derived from the
analysis of the RGS spectra, or to the iron abundance; in this way, the
best-fit model also reproduces sufficiently well the weak spectral feature visible
at the expected position of the \ion{S}{xv} triplet (2.43--2.46\,keV).

\section{Discussion}
\label{sec:discuss}

\subsection{Abundance depletion in the corona of $\tau$~Boo}

\begin{figure} 
\resizebox{\hsize}{!}{\includegraphics[clip,bb=20 440 530 800]{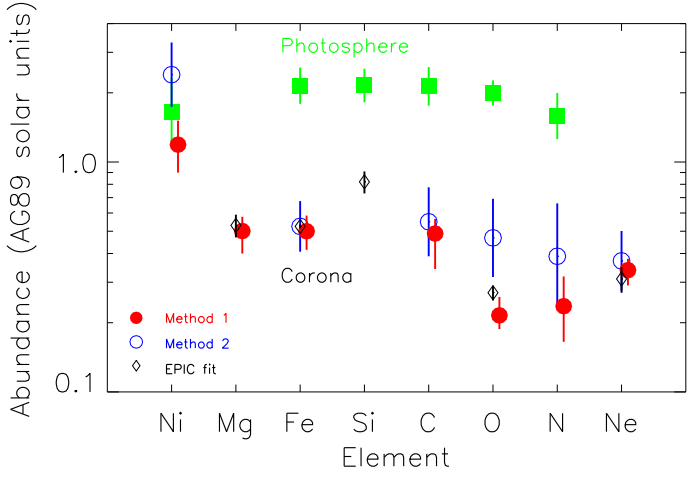}}
\caption{Photospheric and coronal abundances of $\tau$~Boo, sorted by
increasing First Ionization Potential. The reference solar abundances
are by \citet{ag89}.
}
\label{fig:taubooab} 
\end{figure}

The central result of our analysis is illustrated in Fig.\ \ref{fig:taubooab}:
the coronal abundances of our target are all systematically lower than
the photospheric ones, with the only possible exception of the Ni. In this respect,
there is agreement on the abundances derived with any of the methods we employed 
for all elements but oxygen and nickel, within the statistical uncertainties.
The $2\sigma$ difference in the former case can be easily explained by inspection of 
Fig.\ \ref{fig:emd}, where the emission measure derived with Method 1 
in the temperature range $\log T = 6.3$--6.5 is systematically higher than indicated 
by Method 2: 
since much of the \ion{O}{vii} and \ion{O}{viii} line fluxes are produced in
this range, an uncertainty in the emission measure implies
a systematic error in the abundance of this element. 
The case of the Ni abundance is less obvious, but likely because
in APED/ATOMDB V1.3 the Ni atomic model
is less accurate than for other elements, and the Ni lines are in general relatively
faint and hence affected by larger measurement uncertainties.

Regardless of the method, our results show that
the coronal iron abundance is about half the solar photospheric value,
while the stellar photosphere is twice the same value.  
The statistical uncertainty on the ratio of the photospheric to coronal iron abundance 
is very small: $0.24 \pm 0.09$ with Method 1, and $0.23 \pm 0.06$ with Method 2
(uncertainties computed in quadrature), but an additional (systematic) source of uncertainty on
the absolute coronal abundances derived from high-resolution X-ray spectra can be caused by
the determination of the continuum emission level. However, the completeness and accuracy
of atomic databases for the synthesis of multi-temperature model spectra was significantly improved
in the last decade \citep{brickhouse08}, and the results of our RGS data
analysis are confirmed by the spectral fitting of the EPIC data as well.

We conclude that our results are sufficiently robust to state that the X-ray emitting plasma 
of $\tau$~Boo is significantly depleted in metals compared with the photosphere.
Although the determination of the absolute iron abundance in the corona may be affected by an
uncertainty larger than indicated by our analysis, the pattern of abundances scaled by Fe is 
well constrained and the available RGS and EPIC spectra exclude a plasma metallicity matching
the stellar photospheric composition.

\begin{figure} 
\vbox{
\resizebox{\hsize}{!}{\includegraphics[clip,bb=20 440 530 800]{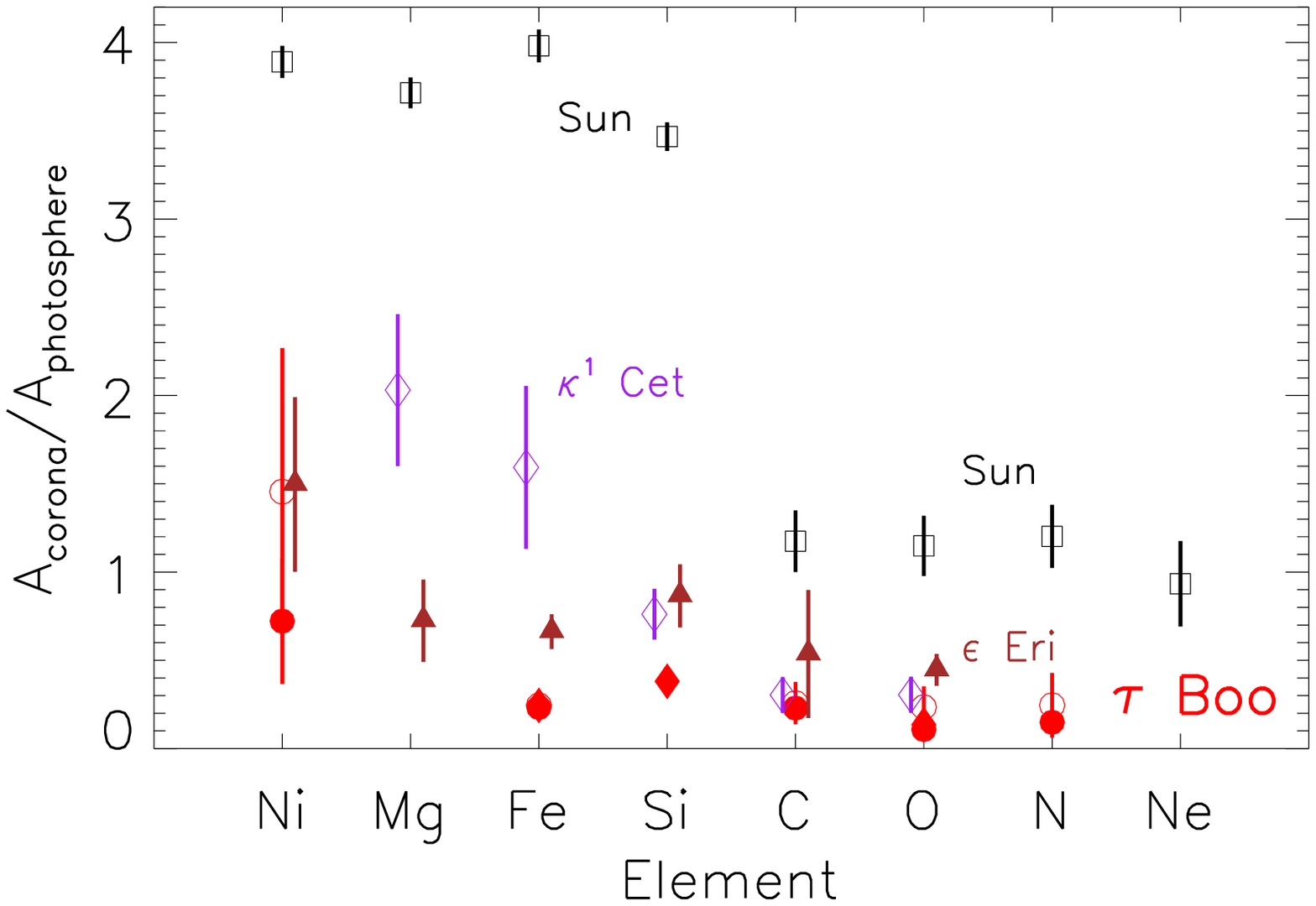}}
\resizebox{\hsize}{!}{\includegraphics[clip,bb=20 440 530 800]{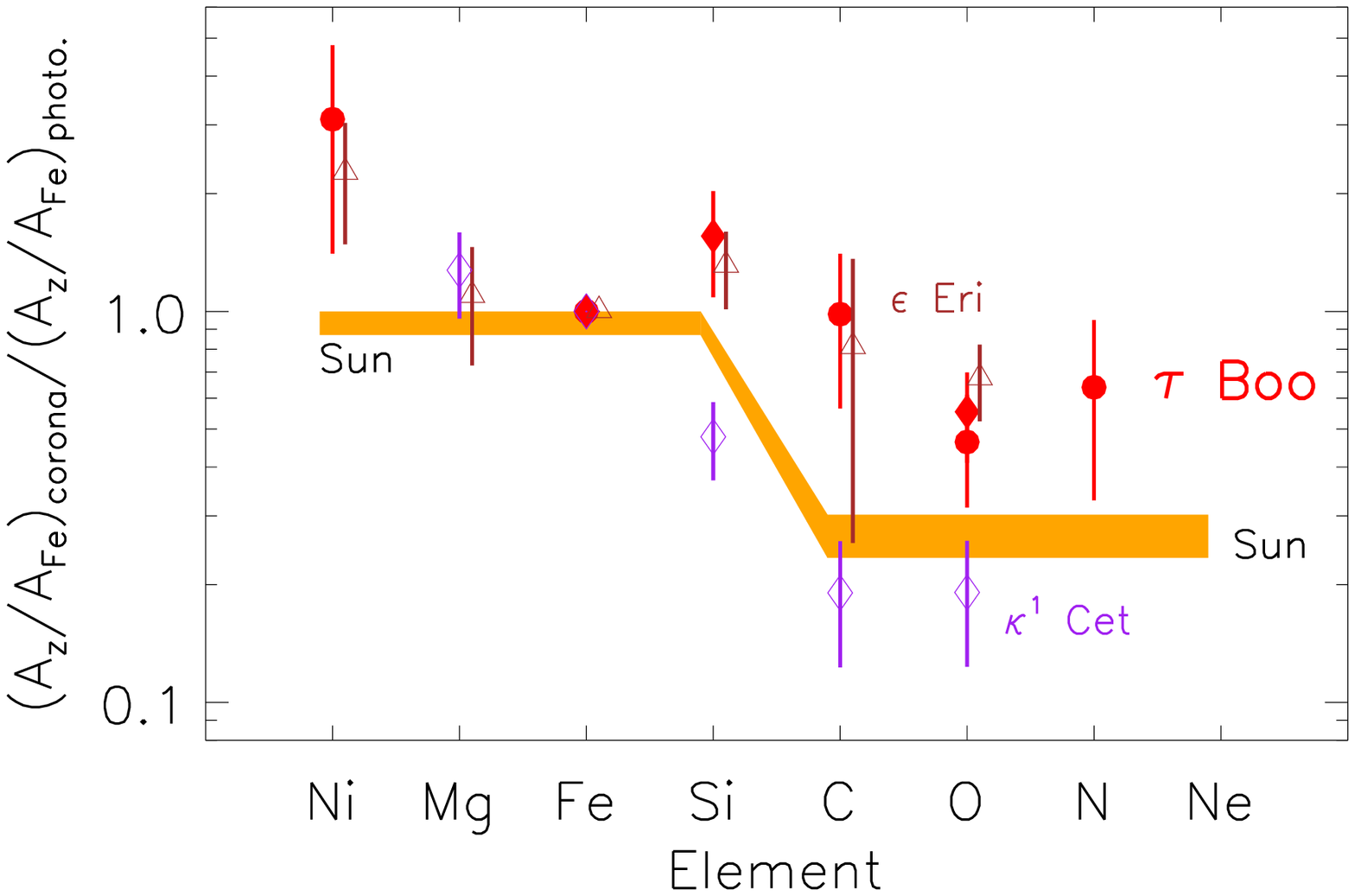}}
}
\caption{{\it Upper:} Coronal to photospheric abundance ratios for the Sun (black squares), 
$\tau$~Boo (red circles and filled diamonds), and two other intermediate-activity stars
(magenta empty diamonds for
$\kappa^1$\,Cet and brown triangles for $\epsilon$\,Eri). Elements are sorted as in Fig.\
\ref{fig:taubooab}. {\it Lower:} Same plot but with all values divided by the Fe
abundance in a logarithmic scale. The FIP bias in the solar corona is indicated by the shaded
(orange) strip, and for $\tau$\,Boo the abundances obtained with Method 2 are not shown 
for clarity.
}
\label{fig:abratios} 
\end{figure}

Moreover, we find
little if any dependence of the coronal abundances on the first ionization potential (FIP),
a result which holds true also considering the coronal vs. photospheric abundance ratios
(Fig.\ \ref{fig:abratios}), rather than values with respect to the solar chemical composition.
In this respect, our results would not change after the possible choice of 
solar element abundances more recent than the \citet{ag89} reference scale, 
which we preferred for consistency with the majority of the previous abundance studies.

\subsection{$\tau$~Boo: a medium-activity coronal source}

\begin{figure} 
\resizebox{\hsize}{!}{\includegraphics{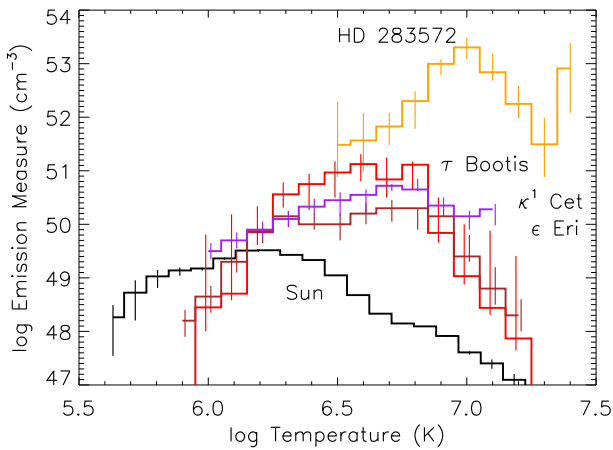}}
\caption{Emission measure distribution of $\tau$~Boo (Method 1)
compared with those of other G- or K-type stars with different activity levels,
including the quiescent solar corona (see text for references).
}
\label{fig:compemd} 
\end{figure}

In order to put the coronal abundances of $\tau$~Boo in context, Fig.\ \ref{fig:compemd}
shows a comparison of its EMD distribution vs.\ temperature (Method 1) with those of other 
four G-K stars:
the Sun \citep[G2V, ][]{peres00}, \object{$\kappa^1$ Ceti} \citep[G5V, ][]{telleschi05}, 
\object{$\epsilon$\,Eri} \citep[K2V, ][]{jsf04}, and \object{HD 283572} \citep[G2IV, ][]{scelsi05}.

The relative amount of plasma at different temperatures is a way to judge the level of
stellar magnetic activity, and it carries more information than a simple comparison of
broad-band X-ray luminosities. Among the stars we selected, the Sun is clearly the
one with the lowest average temperature, with an EMD that peaks just below 2\,MK.
At the other extreme, HD 283572 is a young weak-lined T Tauri star in the Taurus-Auriga
region with a much hotter and X-ray bright corona at 10--20\,MK. The other three stars
were selected instead for the similarity of their EMD with that of $\tau$~Boo.
$\epsilon$\,Eri is also known to host a Jovian-mass planet ($m \sin i \approx 1.6\,{\rm
M_{\rm J}}$) at a distance of 3.4\,AU, while $\kappa^1$\,Cet is a putative single star.

In passing we note that $\tau$\,Boo also has a stellar companion, \object{GJ 527 B}, 
with spectral type M2 (V=11.1) \citep{pwg+02}. 
We evaluated the possible contamination to the observed
X-ray flux by deriving the B-V color and the bolometric correction from the spectral type 
and adopting the transformations by \citet{flower96}, and thus obtaining for the
bolometric luminosity $\log L_{\rm bol} = 30.96~{\rm erg s^{-1}}$. Assuming,
conservatively, that the GJ 527 B X-ray emission is at the saturation level,
we get $L_{\rm x} = 10^{-3} L_{\rm bol} \la 10^{28}~{\rm erg s^{-1}}$, which is about a
factor 10 lower than the X-ray luminosity of $\tau$\,Boo ($\log L_{\rm x} = 28.94~{\rm
erg s^{-1}}$, in the 0.1--2.4\,keV band). We conclude that the contamination from
this unresolved companion is negligible.

We took the photospheric abundances for $\epsilon$\,Eri from
\citet{gonzalezlaws07}, the same source as for $\tau$\,Boo, while for $\kappa^1$\,Cet 
we adopted the photospheric composition by \citet{allendeprieto+04}, which was determined
in a strictly differential way with respect to the Sun. The coronal abundances of the
same two stars were derived by \citet{jsf04} and by
\citet{telleschi05}\footnote{For $\kappa^1$\,Cet we adopted the coronal abundances 
derived with the line-based method and the APEC plasma emissivities in \citet{telleschi05}, 
scaled to the \citet{ag89} solar reference abundances.} respectively,  
with an analysis of XMM-Newton data similar to ours. We then computed the
coronal/photospheric abundance ratios of all the solar-type dwarfs in our sample,
excluding HD 283572 because of its peculiar evolutionary phase and extreme activity
level. 

The pattern of abundance ratios vs. FIP (Fig.\ \ref{fig:abratios}) appears very
similar for $\tau$\,Boo and $\epsilon$\,Eri, and significantly different from the solar
case, which is characterized by the well-known FIP effect \citep{fl00}. 
Finally, $\kappa^1$\,Cet displays an overabundance of low-FIP elements (Mg and Fe) 
in the corona, although with relatively large uncertainties on the abundance ratios, but
high-FIP elements (C and O) are {\it underabundant} with respect to the photospheric
composition, which is not the case in the solar corona. In the right panel of 
Fig.\ \ref{fig:abratios} we also show a comparison of all the coronal/photospheric abundance ratios 
scaled by the iron abundances, because relative abundances are considered more robust than absolute
abundances. This plot clearly shows that in $\tau$\,Boo and $\epsilon$\,Eri there is
a smaller (if any) dependence of the abundance ratios on FIP than in the solar case.

The above comparison suggests that the outer atmospheres of some intermediate 
activity stars, like $\tau$\,Boo and $\epsilon$\,Eri are not affected by a FIP-dependent 
chemical stratification as pronounced as in the solar case. 
A closer look at the left panel in Fig.\ \ref{fig:abratios} shows however that only 
for $\tau$\,Boo the coronal
abundances are significantly and uniformly lower than the photospheric ones (with the
possible exception of Ni).

Common understanding is that the abundance patterns in low- and high-activity
stars are different,
with low-activity stars showing an overabundance of low-FIP elements in the corona
(positive FIP effect), while high-activity stars are characterized by 
the reverse behavior (inverse FIP effect) \citep{ags+03}. 
Following this activity-dependence scenario \citep[see e.g.][]{gn09},
an abundance ratio Ne/Fe $\sim 0.25$ (with respect to the photospheric value) 
is expected for Sun-like stars, while Ne/Fe $\sim 5$--10 in very active stars. 
An intermediate activity star with an EMD similar
to those of $\tau$\,Boo, $\epsilon$\,Eri, and $\kappa^1$\,Cet, should display
Ne/Fe $\sim 1$.

However, this scenario was constructed assuming stellar photospheric abundances
similar to the solar ones -- a common practice for many years -- which has recently been 
challenged by studies where coronal abundances are compared to accurate 
measurements of {\it stellar photospheric abundances}:
in most cases, this properly consistent comparison has shown
little if any significant difference between photospheric and coronal compositions
\citep{jsf04,mff+07,sam09}.

The novel result of our work is that
$\tau$\,Boo does not show any evident FIP bias,
but that on the other hand the coronal abundances are all significantly
lower with respect to the photospheric ones by factors of 3 to 9 (taking into
account the uncertainties, and with the exception of Ni). Hence,
the case of $\tau$\,Boo demonstrates that coronal and photospheric
abundances may differ in a peculiar FIP-independent way, which was never reported 
this clearly before the present study.

This result represents a new challenge for theoretical models that try to explain
element fractionation in stellar outer atmospheres. In spite of many attempts in the
past, the only model that is capable of producing either a positive or an inverse FIP
effect is the one proposed by \citet{l04,l09}. In this model charged ions are subject to a
ponderomotive force caused by Alfv\'en waves propagating through the chromosphere,
which may originate either in the corona or in the convection zone. 
Depending on whether the waves are reflected or not at the base of coronal magnetic
loops, the ponderomotive force can point upward or downward, producing opposite
effects. Low-FIP elements are primarily affected because they are nearly fully ionized in the
upper chromosphere, whereas high-FIP elements are partially neutral. In any case,
the element fractionation is expected to show a FIP-dependence, which is apparently lacking
in the outer atmosphere of $\tau$\,Boo. 

\section{Summary and conclusions}
\label{sec:concl}

We performed a detailed comparison of the photospheric and coronal abundances of the
planet-hosting star $\tau$\,Boo, making use of the best optical and X-ray
spectroscopic data available to date. We analyzed high-resolution spectra obtained with the
reflection gratings (RGS) on-board the XMM-Newton satellite, and also medium-resolution
(EPIC) spectra with much higher photon counting statistics. The emission measure
distribution vs.\ temperature and element abundances of the coronal plasma were
derived with two independent methods from the RGS data, and confirmed with the
analysis of the EPIC spectra.

In spite of some $2\sigma$ difference in the EMDs derived with the two methods,
a robust and consistent result is that the coronal abundances of $\tau$\,Boo 
are significantly lower than the photospheric ones, by about a factor 4, on average. 
The largest uncertainty is for the oxygen abundance, whose underabundance factor 
is in the range 4--9. We do not see any way to bring coronal and photospheric
abundances into agreement, unless the coronal Fe abundance is 4 times
larger than the present best estimate ($\approx 0.5$ the solar one), but such a
systematic error is very unlikely with data of the present quality
and with the state-of-the-art analysis methods we employed.

Assuming solar abundances as reference, the pattern of coronal abundances vs. FIP
is nearly flat, and this behavior is also confirmed when
the actual $\tau$\,Boo photospheric abundances are considered.
No important FIP-dependent element stratification mechanism appears to be at work in the
outer atmosphere of $\tau$\,Boo, but the metallicity in the corona is very different 
from the photospheric one. 
It is unclear at the moment whether this is a peculiarity of metal-rich stars.

From a complementary point of view, while the coronal characteristics
of $\tau$\,Boo including absolute abundances 
are similar to other intermediate-activity stars, 
its photospheric abundances are recognized to be
quite large for Galactic old-disk standards and linked to its being
a star hosting a close-in giant planet. Our results challenge the present understanding
of chemical stratification effects in the atmospheres of solar-type stars, and the
possible role of stellar activity in the underlying physical mechanism.

\begin{acknowledgements}

AM and LS acknowledge partial support for this work from contract
ASI-INAF I/088/06/0 and from an INAF/PRIN grant. 
JSF acknowledge support from the Spanish MICINN through 
grant AYA2008-02038 and the Ram\'on y Cajal Program ref.\ RYC-2005-000549.
Based on observations obtained with {\it XMM-Newton}, an ESA science mission with instruments 
and contributions directly funded by ESA Member States and NASA.

\end{acknowledgements}

\end{document}

%% file: 15587_tab1.tex
\begin{table} 
\begin{minipage}[t]{\columnwidth}
\caption{Adopted photospheric abundances of Tau~Boo}
\label{tab:photab} 
\centering 
\renewcommand{\footnoterule}{}  
\begin{tabular}{cc} 
\hline\hline 
Element & [Ab/H]
\\
\hline 
[C/H] & $0.329 \pm 0.084$ \\
$\mathrm{[N/H]}$ & $0.2 \pm 0.1$ \\ 
$\mathrm{[O/H]}$ & $0.300 \pm 0.055$ \\ 
$\mathrm{[Na/H]}$ & $0.463 \pm 0.096$ \\ 
$\mathrm{[Al/H]}$ & $0.398 \pm 0.100$ \\ 
$\mathrm{[Si/H]}$ & $0.333 \pm 0.073$ \\
$\mathrm{[Ca/H]}$ & $0.161 \pm 0.104$ \\ 
$\mathrm{[Fe/H]}$ & $0.332 \pm 0.079$ \\ 
$\mathrm{[Ni/H]}$ & $0.217 \pm 0.149$ \\ 
\hline 
\end{tabular}
\tablefoot{
All abundances are
from \citet{gonzalezlaws07}, except [N/H] from \citet{ecuvillonetal04b}.
}
\end{minipage}
\end{table} 

%% file: 15587_tab2.tex
\small
\longtab{2}{
\begin{longtable}{lrrrrrrc}
\caption{\label{tab:lines} Line identifications and fluxes in the Tau Boo spectrum.}\\
\hline\hline
 & \multicolumn{3}{c}{Method2} & \multicolumn{3}{c}{Method1} & EMD \\
\cline{2-4} \cline{6-7}
\multicolumn{1}{l}{Ion} & 
\multicolumn{1}{c}{$\lambda_1$} & $f_1 \pm \sigma$ & $(\Delta f/\sigma)^a$ &
\multicolumn{1}{c}{$\lambda_2$} & $f_2 \pm \sigma$ & $(\Delta f/\sigma)^a$ & 
Flag \\
& \multicolumn{1}{c}{\AA} & \multicolumn{2}{l}{$10^{-6}$\,photons cm$^{-2}$\,s$^{-1}$} &
 \multicolumn{1}{c}{\AA} & \multicolumn{2}{l}{$10^{-6}$\,photons cm$^{-2}$\,s$^{-1}$} \\
\hline
\endfirsthead
\caption{continued.}\\
\hline\hline
 & \multicolumn{3}{c}{Method2} & \multicolumn{3}{c}{Method1} & EMD \\
\cline{2-4} \cline{6-7}
\multicolumn{1}{l}{Ion} & 
\multicolumn{1}{c}{$\lambda_1$} & $f_1 \pm \sigma$ & $(\Delta f/\sigma)^a$ & 
\multicolumn{1}{c}{$\lambda_2$} & $f_2 \pm \sigma$ & $(\Delta f/\sigma)^a$ & Flag$^b$ \\
& \multicolumn{1}{c}{\AA} & \multicolumn{2}{l}{$10^{-6}$\,photons cm$^{-2}$\,s$^{-1}$} &
 \multicolumn{1}{c}{\AA} & \multicolumn{2}{l}{$10^{-6}$\,photons cm$^{-2}$\,s$^{-1}$} \\
\hline
\endhead
\hline
\endfoot
     \ion{Mg}{IX} &    9.20 &   \multicolumn{1}{r}{$\cdots$} &    \multicolumn{1}{r}{$\cdots$} &    9.20 &  10.9 $\pm$  1.9 &   -0.0 &  1 \\
   \ion{Fe}{XVII} &   11.25 &  15.2 $\pm$  2.1 &    3.8 &     \multicolumn{1}{r}{$\cdots$} &   \multicolumn{1}{r}{$\cdots$} &    \multicolumn{1}{r}{$\cdots$} &  2 \\
      \ion{Ne}{X} &   12.13 &  28.7 $\pm$  7.0 &   -1.1 &   12.13 &  33.1 $\pm$  4.9 &    0.1 &  3 \\
   \ion{Fe}{XVII} &   12.27 &   9.3 $\pm$  3.1 &   -1.9 &   12.28 &   8.6 $\pm$  3.7 &   -1.5 &  1 \\
    \ion{Ni}{XIX} &   12.42 &   \multicolumn{1}{r}{$\cdots$} &    \multicolumn{1}{r}{$\cdots$} &   12.42 &   7.0 $\pm$  3.5 &   -0.3 &  1 \\
     \ion{Fe}{XX} &   12.57 &   \multicolumn{1}{r}{$\cdots$} &    \multicolumn{1}{r}{$\cdots$} &   12.57 &   8.8 $\pm$  3.5 &    2.4 &   \\
   \ion{Fe}{XVII} &   13.15 &   1.3 $\pm$  0.7 &   -0.4 &     \multicolumn{1}{r}{$\cdots$} &   \multicolumn{1}{r}{$\cdots$} &    \multicolumn{1}{r}{$\cdots$} &   \\
     \ion{Fe}{XX} &   13.27 &   1.3 $\pm$  0.6 &    2.0 &     \multicolumn{1}{r}{$\cdots$} &   \multicolumn{1}{r}{$\cdots$} &    \multicolumn{1}{r}{$\cdots$} &   \\
  \ion{Fe}{XVIII} &   13.32 &   1.3 $\pm$  0.6 &    0.1 &     \multicolumn{1}{r}{$\cdots$} &   \multicolumn{1}{r}{$\cdots$} &    \multicolumn{1}{r}{$\cdots$} &   \\
  \ion{Fe}{XVIII} &   13.39 &   1.3 $\pm$  0.6 &    0.7 &     \multicolumn{1}{r}{$\cdots$} &   \multicolumn{1}{r}{$\cdots$} &    \multicolumn{1}{r}{$\cdots$} &   \\
     \ion{Ne}{IX} &   13.45 &  34.0 $\pm$  4.4 &    0.4 &   13.45 &  22.1 $\pm$  5.0 &   -0.5 &  3 \\
    \ion{Fe}{XIX} &   13.49 &   \multicolumn{1}{r}{$\cdots$} &    \multicolumn{1}{r}{$\cdots$} &   13.49 &  12.7 $\pm$  4.7 &    1.0 &  1 \\
     \ion{Ne}{IX} &   13.55 &   \multicolumn{1}{r}{$\cdots$} &    \multicolumn{1}{r}{$\cdots$} &   13.55 &   0.4 $\pm$  3.7 &   -1.2 &   \\
     \ion{Ne}{IX} &   13.70 &  14.5 $\pm$  2.0 &    0.8 &   13.70 &  13.3 $\pm$  4.0 &    0.2 &  2 \\
\ion{Fe}{XVIII} +  &   13.76 &   \multicolumn{1}{r}{$\cdots$} &    \multicolumn{1}{r}{$\cdots$} &   13.76 &   9.3 $\pm$  2.8 &   -2.7 &   \\
   \ion{Fe}{XVII} &   13.82 &  13.7 $\pm$  2.7 &   -2.7 &   13.83 &   6.9 $\pm$  2.6 &    1.7 &  2 \\
\ion{Fe}{XVIII} +  &   13.94 &   \multicolumn{1}{r}{$\cdots$} &    \multicolumn{1}{r}{$\cdots$} &   13.94 &   9.2 $\pm$  2.5 &    2.9 &   \\
    \ion{Ni}{XIX} &   14.04 &  17.9 $\pm$  2.6 &   -0.1 &   14.06 &  11.6 $\pm$  2.9 &    0.6 &  3 \\
  \ion{Fe}{XVIII} &   14.21 &  19.3 $\pm$  2.4 &   -3.2 &   14.21 &  22.7 $\pm$  3.0 &   -1.2 &  3 \\
  \ion{Fe}{XVIII} &   14.27 &   \multicolumn{1}{r}{$\cdots$} &    \multicolumn{1}{r}{$\cdots$} &   14.27 &   2.5 $\pm$  2.4 &   -1.3 &   \\
  \ion{Fe}{XVIII} &   14.37 &   8.3 $\pm$  1.8 &   -2.4 &   14.37 &  12.3 $\pm$  2.6 &    1.0 &  2 \\
  \ion{Fe}{XVIII} &   14.43 &   0.6 $\pm$  0.3 &   -6.3 &     \multicolumn{1}{r}{$\cdots$} &   \multicolumn{1}{r}{$\cdots$} &    \multicolumn{1}{r}{$\cdots$} &   \\
     \ion{Fe}{XX} &   14.46 &   0.6 $\pm$  0.3 &    0.8 &     \multicolumn{1}{r}{$\cdots$} &   \multicolumn{1}{r}{$\cdots$} &    \multicolumn{1}{r}{$\cdots$} &   \\
  \ion{Fe}{XVIII} &   14.53 &   4.2 $\pm$  1.1 &   -4.0 &   14.52 &   2.5 $\pm$  2.2 &   -1.9 &  2 \\
  \ion{Fe}{XVIII} &   14.58 &   \multicolumn{1}{r}{$\cdots$} &    \multicolumn{1}{r}{$\cdots$} &   14.58 &   9.4 $\pm$  2.3 &    2.2 &   \\
\ion{Fe}{XVIII} + &   14.76 &   \multicolumn{1}{r}{$\cdots$} &    \multicolumn{1}{r}{$\cdots$} &   14.76 &   6.3 $\pm$  2.2 &    2.2 &   \\
    \ion{O}{VIII} &   14.82 &   0.8 $\pm$  0.3 &  -11.4 &     \multicolumn{1}{r}{$\cdots$} &   \multicolumn{1}{r}{$\cdots$} &    \multicolumn{1}{r}{$\cdots$} &   \\
     \ion{Fe}{XX} &   14.92 &   1.7 $\pm$  0.5 &    3.0 &     \multicolumn{1}{r}{$\cdots$} &   \multicolumn{1}{r}{$\cdots$} &    \multicolumn{1}{r}{$\cdots$} &  2 \\
   \ion{Fe}{XVII} &   15.01 & 111.9 $\pm$  4.0 &  -12.3 &   15.02 & 126.6 $\pm$  5.3 &   -4.1 &  3 \\
    \ion{Fe}{XIX} &   15.08 &   3.9 $\pm$  1.8 &    1.0 &     \multicolumn{1}{r}{$\cdots$} &   \multicolumn{1}{r}{$\cdots$} &    \multicolumn{1}{r}{$\cdots$} &   \\
    \ion{O}{VIII} &   15.18 &  21.2 $\pm$  3.2 &    2.8 &   15.17 &  20.5 $\pm$  3.6 &    3.3 &  2 \\
   \ion{Fe}{XVII} &   15.26 &  42.1 $\pm$  3.1 &   -1.9 &   15.27 &  48.3 $\pm$  4.0 &   -0.1 &  3 \\
   \ion{Fe}{XVIII} &   15.40 &   \multicolumn{1}{r}{$\cdots$} &    \multicolumn{1}{r}{$\cdots$} &   15.40 &   4.1 $\pm$  2.8 &    0.3 &   \\
   \ion{Fe}{XVII} &   15.45 &  28.3 $\pm$  3.6 &    5.1 &   15.46 &  23.1 $\pm$  3.0 &    5.3 &  2 \\
  \ion{Fe}{XVIII} &   15.62 &   9.9 $\pm$  1.8 &    2.1 &   15.61 &  13.3 $\pm$  2.5 &    2.3 &  3 \\
  \ion{Fe}{XVIII} &   15.82 &   6.9 $\pm$  1.0 &    1.1 &     \multicolumn{1}{r}{$\cdots$} &   \multicolumn{1}{r}{$\cdots$} &    \multicolumn{1}{r}{$\cdots$} &  2 \\
  \ion{Fe}{XVIII} &   15.87 &   \multicolumn{1}{r}{$\cdots$} &    \multicolumn{1}{r}{$\cdots$} &   15.87 &   9.6 $\pm$  2.2 &    2.6 &   \\
    \ion{O}{VIII} &   16.01 &  28.1 $\pm$  4.3 &   -2.2 &   16.00 &  30.6 $\pm$  3.3 &    1.5 &  2 \\
  \ion{Fe}{XVIII} &   16.07 &  15.9 $\pm$  2.9 &    0.3 &   16.09 &  17.8 $\pm$  2.9 &    1.7 &  3 \\
  \ion{Fe}{XVIII} &   16.16 &   \multicolumn{1}{r}{$\cdots$} &    \multicolumn{1}{r}{$\cdots$} &   16.16 &   1.2 $\pm$  2.2 &   -1.2 &   \\
\ion{Fe}{XIX} + &   16.29 &   \multicolumn{1}{r}{$\cdots$} &    \multicolumn{1}{r}{$\cdots$} &   16.29 &   3.2 $\pm$  2.3 &    1.0 &   \\
\ion{Fe}{XVIII} +  &   16.37 &   \multicolumn{1}{r}{$\cdots$} &    \multicolumn{1}{r}{$\cdots$} &   16.37 &   3.4 $\pm$  2.6 &    0.0 &   \\
   \ion{Fe}{XVII} &   16.78 &  78.2 $\pm$  5.1 &    0.9 &   16.78 &  85.0 $\pm$  4.5 &    3.4 &  3 \\
   \ion{Fe}{XVII} &   17.05 & 197.4 $\pm$ 11.9 &    2.2 &   17.06 & 108.4 $\pm$  5.3 &    4.9 &  3 \\
   \ion{Fe}{XVIII} &   17.09 &   \multicolumn{1}{r}{$\cdots$} &    \multicolumn{1}{r}{$\cdots$} &   17.09 &  89.0 $\pm$  4.4 &    2.4 &   \\
  \ion{Fe}{XVIII} &   17.62 &   5.8 $\pm$  2.0 &   -0.3 &     \multicolumn{1}{r}{$\cdots$} &   \multicolumn{1}{r}{$\cdots$} &    \multicolumn{1}{r}{$\cdots$} &   \\
     \ion{O}{VII} &   18.63 &   6.6 $\pm$  2.0 &   -0.3 &     \multicolumn{1}{r}{$\cdots$} &   \multicolumn{1}{r}{$\cdots$} &    \multicolumn{1}{r}{$\cdots$} &  2 \\
  \ion{\ion{C}a}{XVIII} &   18.69 &   0.3 $\pm$  0.2 &    0.5 &     \multicolumn{1}{r}{$\cdots$} &   \multicolumn{1}{r}{$\cdots$} &    \multicolumn{1}{r}{$\cdots$} &   \\
    \ion{O}{VIII} &   18.97 & 126.0 $\pm$  6.5 &  -12.7 &   18.97 & 129.0 $\pm$  5.9 &    0.2 &  3 \\
    \ion{\ion{C}a}{X}VI &   21.45 &   0.6 $\pm$  0.5 &   -1.1 &     \multicolumn{1}{r}{$\cdots$} &   \multicolumn{1}{r}{$\cdots$} &    \multicolumn{1}{r}{$\cdots$} &   \\
     \ion{O}{VII} &   21.60 &  48.9 $\pm$  7.4 &   -0.5 &   21.60 &  53.8 $\pm$  7.2 &   -0.7 &  3 \\
     \ion{O}{VII} &   21.80 &   9.0 $\pm$  3.8 &    0.6 &   21.83 &   8.9 $\pm$  4.6 &    0.2 &   \\
     \ion{O}{VII} &   22.10 &  41.3 $\pm$  7.1 &    2.0 &   22.10 &  43.0 $\pm$  6.9 &    1.4 &  3 \\
     N{VII} &   24.78 &  16.0 $\pm$  3.6 &   -0.0 &   24.80 &  15.6 $\pm$  3.9 &    0.3 &  3 \\
    \ion{Ar}{X}VI &   24.99 &   5.3 $\pm$  1.9 &    0.0 &     \multicolumn{1}{r}{$\cdots$} &   \multicolumn{1}{r}{$\cdots$} &    \multicolumn{1}{r}{$\cdots$} &   \\
      \ion{C}{VI} &   28.47 &  10.1 $\pm$  2.5 &    0.1 &     \multicolumn{1}{r}{$\cdots$} &   \multicolumn{1}{r}{$\cdots$} &    \multicolumn{1}{r}{$\cdots$} &  2 \\
      \ion{C}{VI} &   33.73 &  42.8 $\pm$  6.6 &   -0.1 &   33.74 &  58.9 $\pm$  7.9 &   -0.2 &  3 \\
\end{longtable}
\tablefoot{
\tablefoottext{a}
Observed minus predicted line flux, divided by flux error.
\tablefoottext{b}
Flag=1 if line was employed for EMD reconstruction with Method 1 only,
Flag=2 if Method 2 only, Flag=3 if employed by both methods.
} 
}
\normalsize

%% file: 15587_tab3.tex
\begin{table} 
\begin{minipage}[t]{\columnwidth}
\caption{Volume emission measure distribution and coronal abundances of Tau~Boo}
\label{tab:emd} 
\centering 
\begin{tabular}{cccc} 
\hline\hline 
        & \multicolumn{2}{c}{RGS} \\
        & Method 1 & Method 2 & EPIC \\
\hline 
$\log T$\,[K] & \multicolumn{2}{c}{$\log EM\,\mathrm{[cm^{-3}]}$} \\
\hline 
 6.0 & 48.45$^{1.36}_{0.45}$ & 49.28                 & ...  \\
 6.1 & 48.70$^{1.48}_{0.12}$ & 49.38                 & ...  \\
 6.2 & 49.87$^{0.46}_{0.26}$ & 49.58$^{0.20}_{0.40}$ & ...  \\
 6.3 & 50.56$^{0.22}_{0.25}$ & 49.88$^{0.20}_{0.30}$ & 50.94$^{0.08}_{0.05}$ \\
 6.4 & 50.75$^{0.19}_{0.48}$ & 50.08$^{0.20}_{0.20}$ & ...  \\
 6.5 & 50.97$^{0.21}_{0.47}$ & 50.38$^{0.40}_{0.20}$ & ...  \\
 6.6 & 51.13$^{0.18}_{0.33}$ & 51.23$^{0.20}_{0.05}$ & 51.70$^{0.02}_{0.02}$ \\
 6.7 & 50.84$^{0.41}_{0.34}$ & 51.43$^{0.05}_{0.20}$ & ...  \\
 6.8 & 51.11$^{0.05}_{0.47}$ & 50.58$^{0.00}_{0.20}$ & ...  \\
 6.9 & 49.84$^{0.66}_{0.18}$ & 49.78$^{0.20}_{0.40}$ & 50.93$^{0.06}_{0.04}$ \\
 7.0 & 49.03$^{0.97}_{0.16}$ & 49.58$^{0.20}_{0.30}$ & ...  \\
 7.1 & 48.44$^{1.44}_{0.13}$ & 49.48$^{0.30}_{0.40}$ & ...  \\
 7.2 & 47.87$^{1.54}_{0.22}$ & 49.28$^{0.30}_{0.30}$ & ...  \\
\hline 
Elem. & \multicolumn{2}{c}{$A_{\rm c}/A_{\odot}$\tablefootmark{a}} &
$A_{\rm c}/A_{\odot}$\tablefootmark{a} \\
\hline 
  C & 0.49$^{0.08}_{0.15}$ & 0.55$^{0.23}_{0.16}$ & 0.6 \\
  N & 0.24$^{0.08}_{0.07}$ & 0.39$^{0.27}_{0.16}$ & 0.4 \\
  O & 0.22$^{0.04}_{0.03}$ & 0.47$^{0.22}_{0.15}$ & 0.27$^{0.02}_{0.01}$ \\
 Ne & 0.34$^{0.04}_{0.05}$ & 0.37$^{0.13}_{0.10}$ & 0.31$^{0.04}_{0.04}$ \\
 Mg & 0.50$^{0.08}_{0.10}$ &          ...         & 0.53$^{0.06}_{0.04}$ \\
 Si &        ...           &          ...         & 0.82$^{0.09}_{0.08}$ \\
  S &        ...           &          ...         & = Fe \\
 Ar &        ...           &          ...         & = Fe \\
 Fe & 0.50$^{0.08}_{0.09}$ & 0.52$^{0.15}_{0.12}$ & 0.52$^{0.01}_{0.01}$ \\
 Ni & 1.19$^{0.32}_{0.29}$ & 2.40$^{0.91}_{0.66}$ & = Fe \\
\hline 
\end{tabular}
\tablefoot{
\tablefoottext{a}{All abundances are relative to the solar photospheric values 
by \citet{ag89}. Values with no error were fixed (see text).}
}
\end{minipage}
\end{table}